\documentclass[12pt,reqno]{amsart}
\usepackage{graphicx}
\usepackage{amscd}
\usepackage{amsmath}
\usepackage{epsfig}
\usepackage{amsfonts}
\usepackage{amssymb}

\providecommand{\U}[1]{\protect\rule{.1in}{.1in}}
\providecommand{\U}[1]{\protect\rule{.1in}{.1in}}
\allowdisplaybreaks

\theoremstyle{plain}

\newtheorem{lemma}{Lemma}

\numberwithin{equation}{section}

\input{tcilatex}

\begin{document}
\title[ Nonlinear Schr\"{o}dinger Equations]{On Integrability of
Nonautonomous Nonlinear Schr\"{o}dinger Equations}
\author{Sergei K. Suslov}
\address{School of Mathematical and Statistical Sciences \& Mathematical,
Computational and Modeling Sciences Center, Arizona State University, Tempe,
AZ 85287--1804, U.S.A.}
\email{sks@asu.edu}
\urladdr{http://hahn.la.asu.edu/\symbol{126}suslov/index.html}
\date{\today }
\subjclass{Primary 35Q55, 35Q51. Secondary 35P30, 81Q05}
\keywords{Nonlinear Schr\"{o}dinger equations, generalized harmonic
oscillators, Green's function, propagator, completely integrable systems,
Lax pair, Zakharov--Shabat system.}

\begin{abstract}
We show, in general, how to transform the nonautonomous nonlinear Schr\"{o}%
dinger equation with quadratic Hamiltonians into the standard autonomous
form that is completely integrable by the familiar inverse scattering method
in nonlinear science. Derivation of the corresponding equivalent
nonisospectral Lax pair is also outlined. A few simple integrable systems
are discussed.
\end{abstract}

\maketitle

\section{Introduction}

Recently several nonautonomous (with time-dependent coefficients) and
inhomogeneous (with space-dependent coefficients) nonlinear Schr\"{o}dinger
equations have been discussed as (possible) new integrable systems \cite%
{AlKha10}, \cite{Atre:Pani:Aga06}, \cite{Bol-BPerezVersKonotop08}, \cite%
{Bruga:Sci10}, \cite{ChenYi05}, \cite{Eba:Khal}, \cite{HeLi10}, \cite%
{Kruglovetal03}, \cite{Krugloveta05}, \cite{LiangZhangLiu05}, \cite{Moores96}%
, \cite{Moores01}, \cite{PonomAgr07}, \cite{Serkin:Hasrgawa00}, \cite%
{Serkin:Hasegawa00}, \cite{Serkinetal04}, \cite{Serkinetal07}, \cite%
{Serkinetal10}, \cite{SuazoSuslovSol}, \cite%
{Trall-Gin:Drake:Lop-Rich:Trall-Herr:Bir}, \cite{ZYan03a}, \cite{Zyan04},
\cite{ZYan10}, \cite{Zhangetal08} (see also \cite{AblowClark91}, \cite%
{AblowPrinTrub04}, \cite{Balakrish85}, \cite{ChenHH:LiuCS76}, \cite%
{ChenHH:LiuCS78},~\cite{Clark88}, \cite{GagWint93}, \cite{Kundu09}, \cite%
{TappZab71} and references therein for earlier works). They arise in the
theory of Bose--Einstein condensation \cite{Dal:Giorg:Pitaevski:Str99}, \cite%
{Pit:StrinBook}, fiber optics \cite{Agrawal}, \cite{Hasegawa},
superconductivity and plasma physics \cite{ChenHH:LiuCS76}, \cite%
{ChenHH:LiuCS78}, \cite{Newell78}, \cite{Novikovetal}.

As pointed out in recent papers~\cite{AlKha10}, \cite{Bruga:Sci10}, \cite%
{HeLi10}, \cite{Heetal09} and \cite{Kundu09} (see also \cite{AblowClark91},
\cite{AblowPrinTrub04}, \cite{Clark88}, \cite{Cosgrove93II}, \cite{GagWint93}%
, \cite{Musette99}, \cite{Per-GTorrKonot06}), these systems can be reduced
by a set of transformations to the standard autonomous nonlinear Schr\"{o}%
dinger equation, which explains their integrability properties because this
equation is a well-known complete integrable system with Lax pair \cite%
{Lax68}, \cite{Zakh:Shab71}, \cite{ZakhShab74}, \cite{ZakhShab79},
conservation laws and $N$-soliton solutions, solvable through the inverse
scattering method \cite{AblowClark91}, \cite{AblowPrinTrub04}, \cite%
{Ablo:Seg81}, \cite{KudryashovBook10}, \cite{Novikovetal}, \cite{SatYaj74},
\cite{Zakh:Shab71}, \cite{ZakhShab74}, \cite{ZakhShab79}. Integration
techniques of the nonlinear Schr\"{o}dinger equation include also Painlev%
\'{e} analysis \cite{BoitiPem80}, \cite{Conte89}, \cite{Conte99}, \cite%
{ConteFordyPick93}, \cite{ConteMusette09}, \cite{Cosgrove93I}, \cite%
{Cosgrove93II}, \cite{Heetal09}, \cite{KudryashovBook10}, \cite{MusConte03},
\cite{Tajiri83}, \cite{Weissetal82}, Hirota method \cite{Hirota71}, \cite%
{HirotaBook}, \cite{KudryashovBook10}, B\"{a}cklund transformation \cite%
{BoitiGuiz82}, \cite{ChenHH74}, \cite{KudryashovBook10} and Hamiltonian
approach \cite{Ablowitzetal73}, \cite{Ablo:Seg81}, \cite{Fadd:Takh}, \cite%
{Gardneretal67}, \cite{Miura68}, \cite{Miuraetal68}, \cite{Novikovetal}
among others \cite{DegasRes}, \cite{Mateev:SalleBook}, \cite{OlverPBook},
\cite{Rogers:SchiefBook}.

A goal of this paper is to construct these transformations explicitly (in
quadratures) for the most general variable quadratic Hamiltonian. A simple
relation with Green's function of the linear problem, which seems has been
missing in the available literature, is emphasized. Basics of the classical
soliton theory including one and two soliton solutions, inverse scattering
technique and the corresponding equivalent Lax pair are also briefly
summarized in order to make our presentation as self-contained as possible.
It may facilitate applications of our result to specific nonlinear
integrable systems.

\section{Transformation into Autonomous Form}

We consider the nonautonomous nonlinear Schr\"{o}dinger equation:%
\begin{equation}
i\frac{\partial \psi }{\partial t}=H\psi +h\left\vert \psi \right\vert
^{2}\psi  \label{SolInt1}
\end{equation}%
on $%
\mathbb{R}
,$ where the variable Hamiltonian $H=Q\left( p,x,t\right) $ is an arbitrary
quadratic form of operators $p=-i\partial /\partial x$ and $x,$ namely,%
\begin{equation}
i\psi _{t}=-a\left( t\right) \psi _{xx}+b\left( t\right) x^{2}\psi -ic\left(
t\right) x\psi _{x}-id\left( t\right) \psi -f\left( t\right) x\psi +ig\left(
t\right) \psi _{x}+h\left( t\right) \left\vert \psi \right\vert ^{2}\psi ,
\label{SolInt2}
\end{equation}%
($a,$ $b,$ $c,$ $d,$ $f$ and $g$ are suitable real-valued functions of time
only) under the following integrability condition\footnote{%
If the nonlinear term has the form $h\left\vert \psi \right\vert ^{p}\psi ,$
popular in the mathematical literature, the condition becomes $h=h_{0}a\beta
^{2}\mu ^{p/2}.$} \cite{SuazoSuslovSol}:%
\begin{equation}
h=h_{0}a\left( t\right) \beta ^{2}\left( t\right) \mu \left( t\right)
=h_{0}\beta ^{2}\left( 0\right) \mu ^{2}\left( 0\right) \frac{a\left(
t\right) \lambda ^{2}\left( t\right) }{\mu \left( t\right) }
\label{SolInt2a}
\end{equation}%
($h_{0}$ is a real constant, functions $\beta ,$ $\lambda $ and $\mu $ will
be defined below).

We present the following result.

\begin{lemma}
The substitution%
\begin{equation}
\psi \left( x,t\right) =\frac{1}{\sqrt{\mu \left( t\right) }}e^{i\left(
\alpha \left( t\right) x^{2}+\delta \left( t\right) x+\kappa \left( t\right)
\right) }\ \chi \left( \xi ,\tau \right) ,\qquad \xi =\beta \left( t\right)
x+\varepsilon \left( t\right) ,\quad \tau =\gamma \left( t\right)
\label{SolIntTransform}
\end{equation}%
transforms the nonautonomous and inhomogeneous nonlinear Schr\"{o}dinger
equation (\ref{SolInt2}) into the standard autonomous form with respect to
new variables $\xi =\beta \left( t\right) x+\varepsilon \left( t\right) $
and $\tau =\gamma \left( t\right) $ $:$%
\begin{equation}
i\chi _{\tau }+h_{0}\left\vert \chi \right\vert ^{2}\chi =\chi _{\xi \xi }
\label{SolInt3}
\end{equation}%
provided that%
\begin{equation}
\frac{d\alpha }{dt}+b+2c\alpha +4a\alpha ^{2}=0,  \label{SysA}
\end{equation}%
\begin{equation}
\frac{d\beta }{dt}+\left( c+4a\alpha \right) \beta =0,  \label{SysB}
\end{equation}%
\begin{equation}
\frac{d\gamma }{dt}+a\beta ^{2}=0  \label{SysC}
\end{equation}%
and%
\begin{equation}
\frac{d\delta }{dt}+\left( c+4a\alpha \right) \delta =f+2\alpha g,
\label{SysD}
\end{equation}%
\begin{equation}
\frac{d\varepsilon }{dt}=\left( g-2a\delta \right) \beta ,  \label{SysE}
\end{equation}%
\begin{equation}
\frac{d\kappa }{dt}=g\delta -a\delta ^{2},  \label{SysF}
\end{equation}%
where%
\begin{equation}
\alpha =\frac{1}{4a\left( t\right) }\frac{\mu ^{\prime }\left( t\right) }{%
\mu \left( t\right) }-\frac{d\left( t\right) }{2a\left( t\right) }.
\label{Alpha}
\end{equation}
\end{lemma}

The autonomous equation (\ref{SolInt3}) is completely integrable by advanced
methods of the soliton theory \cite{Ablo:Seg81}, \cite{KudryashovBook10},
\cite{Novikovetal}, \cite{Scott03}, \cite{Zakh:Shab71}, \cite{ZakhShab74},
\cite{ZakhShab79} (see also \cite{DegasRes} and references cited in the
introduction). Equations (\ref{SolInt2})--(\ref{SolInt2a}) seem to represent
the maximum nonautonomous and inhomogeneous one-dimensional integrable
system of this kind. (Important special cases of the transformation (\ref%
{SolIntTransform}) are discussed in Refs.~\cite{AblowClark91}, \cite%
{ChenHH:LiuCS76}, \cite{ChenHH:LiuCS78}, \cite{Clark88}, \cite{GagWint93},
\cite{Heetal09}, \cite{Kundu09}, \cite{Per-GTorrKonot06}, \cite{PonomAgr07}
and \cite{SuazoSuslovSol}.)

Our transformation (\ref{SolIntTransform}) involves the real-valued
functions $\alpha ,$ $\beta ,$ $\gamma ,$ $\delta ,$ $\varepsilon $ and $%
\kappa $ (of time $t$ only) defined as solutions of the system of ordinary
differential equations (\ref{SysA})--(\ref{SysF}). This nonlinear ODE system
has already appeared in Ref.~\cite{Cor-Sot:Lop:Sua:Sus} from a different
perspective (we shall refer to this system as a Riccati-type system). The
substitution (\ref{Alpha}) reduces the Riccati equation (\ref{SysA}) to the
second order linear equation \cite{Cor-Sot:Lop:Sua:Sus}:%
\begin{equation}
\mu ^{\prime \prime }-\tau \left( t\right) \mu ^{\prime }+4\sigma \left(
t\right) \mu =0  \label{CharEq}
\end{equation}%
with%
\begin{equation}
\tau \left( t\right) =\frac{a^{\prime }}{a}-2c+4d,\qquad \sigma \left(
t\right) =ab-cd+d^{2}+\frac{d}{2}\left( \frac{a^{\prime }}{a}-\frac{%
d^{\prime }}{d}\right) .  \label{TauSigma}
\end{equation}%
(Relations with the corresponding Ehrenfest theorem for the linear
Hamiltonian are discussed in Ref.~\cite{Cor-Sot:Sua:SusInv}. It provides a
clear physical interpretation of our results in case of the Gross-Pitaevskii
model of Bose condensation.) Equation (\ref{SysC}) implies the monotonicity
of the new time variable $\tau =\gamma \left( t\right) .$

\begin{proof}
Differentiate $\psi =\mu ^{-1/2}\left( t\right) e^{iS\left( x,t\right) }\chi
\left( \xi ,\tau \right) $ with $S=\alpha \left( t\right) x^{2}+\delta
\left( t\right) x+\kappa \left( t\right) $ and $\xi =\beta \left( t\right)
x+\varepsilon \left( t\right) ,$ $\tau =\gamma \left( t\right) :$%
\begin{equation}
ie^{-iS}\psi _{t}=\frac{1}{\sqrt{\mu }}\left[ -\left( \alpha ^{\prime
}x^{2}+\delta ^{\prime }x+\kappa ^{\prime }\right) \chi +i\left( \left(
\beta ^{\prime }x+\varepsilon ^{\prime }\right) \chi _{\xi }+\gamma ^{\prime
}\chi _{\tau }-\frac{\mu ^{\prime }}{2\mu }\chi \right) \right] ,
\label{Pf1}
\end{equation}%
\begin{equation}
e^{-iS}\psi _{x}=\frac{1}{\sqrt{\mu }}\left[ i\left( 2\alpha x+\delta
\right) \chi +\beta \chi _{\xi }\right]  \label{Pf2}
\end{equation}%
and%
\begin{equation}
e^{-iS}\psi _{xx}=\frac{1}{\sqrt{\mu }}\left[ \left( 2i\alpha -\left(
2\alpha x+\delta \right) ^{2}\right) \chi +2i\left( 2\alpha x+\delta \right)
\beta \chi _{\xi }+\beta ^{2}\chi _{\xi \xi }\right] .  \label{Pf3}
\end{equation}%
Substitution into (\ref{SolInt2}), with the help of the integrability
condition (\ref{SolInt2a}) and the system (\ref{SysA})--(\ref{SysF}),
results in (\ref{SolInt3}). Computational details are left to the reader.
\end{proof}

This observation provides a new interpretation of the Riccati-type system (%
\ref{SysA})--(\ref{SysF}), which has been originally derived in Ref.~\cite%
{Cor-Sot:Lop:Sua:Sus} during integration of the corresponding linear
equation via Green's function.

\section{Integration of the Riccati-type System}

In order to construct the transformation (\ref{SolIntTransform}) explicitly,
one has to solve the nonlinear ODE system (\ref{SysA})--(\ref{SysF}). As
already known, the initial value problem of the Riccati-type system, which
corresponds to the linear Schr\"{o}dinger equation with a variable quadratic
Hamiltonian (generalized harmonic oscillators \cite{Berry85}, \cite%
{Dod:Mal:Man75}, \cite{Hannay85}, \cite{Wolf81}, \cite%
{Yeon:Lee:Um:George:Pandey93}), can be explicitly solved in terms of
solutions of characteristic equation (\ref{CharEq}) as follows \cite%
{Cor-Sot:Lop:Sua:Sus}, \cite{Cor-Sot:Sua:SusInv}, \cite{Suaz:Sus}, \cite%
{Suslov10}:%
\begin{eqnarray}
&&\mu \left( t\right) =2\mu \left( 0\right) \mu _{0}\left( t\right) \left(
\alpha \left( 0\right) +\gamma _{0}\left( t\right) \right) ,  \label{MKernel}
\\
&&\alpha \left( t\right) =\alpha _{0}\left( t\right) -\frac{\beta
_{0}^{2}\left( t\right) }{4\left( \alpha \left( 0\right) +\gamma _{0}\left(
t\right) \right) },  \label{AKernel} \\
&&\beta \left( t\right) =-\frac{\beta \left( 0\right) \beta _{0}\left(
t\right) }{2\left( \alpha \left( 0\right) +\gamma _{0}\left( t\right)
\right) }=\frac{\beta \left( 0\right) \mu \left( 0\right) }{\mu \left(
t\right) }\lambda \left( t\right) ,  \label{BKernel} \\
&&\gamma \left( t\right) =\gamma \left( 0\right) -\frac{\beta ^{2}\left(
0\right) }{4\left( \alpha \left( 0\right) +\gamma _{0}\left( t\right)
\right) }  \label{CKernel}
\end{eqnarray}%
and%
\begin{eqnarray}
\delta \left( t\right) &=&\delta _{0}\left( t\right) -\frac{\beta _{0}\left(
t\right) \left( \delta \left( 0\right) +\varepsilon _{0}\left( t\right)
\right) }{2\left( \alpha \left( 0\right) +\gamma _{0}\left( t\right) \right)
},  \label{DKernel} \\
\varepsilon \left( t\right) &=&\varepsilon \left( 0\right) -\frac{\beta
\left( 0\right) \left( \delta \left( 0\right) +\varepsilon _{0}\left(
t\right) \right) }{2\left( \alpha \left( 0\right) +\gamma _{0}\left(
t\right) \right) },  \label{EKernel} \\
\kappa \left( t\right) &=&\kappa \left( 0\right) +\kappa _{0}\left( t\right)
-\frac{\left( \delta \left( 0\right) +\varepsilon _{0}\left( t\right)
\right) ^{2}}{4\left( \alpha \left( 0\right) +\gamma _{0}\left( t\right)
\right) }.  \label{FKernel}
\end{eqnarray}%
Here,%
\begin{eqnarray}
&&\alpha _{0}\left( t\right) =\frac{1}{4a\left( t\right) }\frac{\mu
_{0}^{\prime }\left( t\right) }{\mu _{0}\left( t\right) }-\frac{d\left(
t\right) }{2a\left( t\right) },  \label{A0} \\
&&\beta _{0}\left( t\right) =-\frac{\lambda \left( t\right) }{\mu _{0}\left(
t\right) },\qquad \lambda \left( t\right) =\exp \left( -\int_{0}^{t}\left(
c\left( s\right) -2d\left( s\right) \right) \ ds\right) ,  \label{B0} \\
&&\gamma _{0}\left( t\right) =\frac{1}{2\mu _{1}\left( 0\right) }\frac{\mu
_{1}\left( t\right) }{\mu _{0}\left( t\right) }+\frac{d\left( 0\right) }{%
2a\left( 0\right) }  \label{C0}
\end{eqnarray}%
and%
\begin{equation}
\delta _{0}\left( t\right) =\frac{\lambda \left( t\right) }{\mu _{0}\left(
t\right) }\int_{0}^{t}\left[ \left( f\left( s\right) -\frac{d\left( s\right)
}{a\left( s\right) }g\left( s\right) \right) \mu _{0}\left( s\right) +\frac{%
g\left( s\right) }{2a\left( s\right) }\mu _{0}^{\prime }\left( s\right) %
\right] \frac{ds}{\lambda \left( s\right) },  \label{D0}
\end{equation}%
\begin{eqnarray}
\varepsilon _{0}\left( t\right) &=&-\frac{2a\left( t\right) \lambda \left(
t\right) }{\mu _{0}^{\prime }\left( t\right) }\delta _{0}\left( t\right)
+8\int_{0}^{t}\frac{a\left( s\right) \sigma \left( s\right) \lambda \left(
s\right) }{\left( \mu _{0}^{\prime }\left( s\right) \right) ^{2}}\left( \mu
_{0}\left( s\right) \delta _{0}\left( s\right) \right) \ ds  \label{E0} \\
&&\quad +2\int_{0}^{t}\frac{a\left( s\right) \lambda \left( s\right) }{\mu
_{0}^{\prime }\left( s\right) }\left( f\left( s\right) -\frac{d\left(
s\right) }{a\left( s\right) }g\left( s\right) \right) \ ds,  \notag
\end{eqnarray}%
\begin{eqnarray}
\kappa _{0}\left( t\right) &=&\frac{a\left( t\right) \mu _{0}\left( t\right)
}{\mu _{0}^{\prime }\left( t\right) }\delta _{0}^{2}\left( t\right)
-4\int_{0}^{t}\frac{a\left( s\right) \sigma \left( s\right) }{\left( \mu
_{0}^{\prime }\left( s\right) \right) ^{2}}\left( \mu _{0}\left( s\right)
\delta _{0}\left( s\right) \right) ^{2}\ ds  \label{F0} \\
&&\quad -2\int_{0}^{t}\frac{a\left( s\right) }{\mu _{0}^{\prime }\left(
s\right) }\left( \mu _{0}\left( s\right) \delta _{0}\left( s\right) \right)
\left( f\left( s\right) -\frac{d\left( s\right) }{a\left( s\right) }g\left(
s\right) \right) \ ds  \notag
\end{eqnarray}%
$(\delta _{0}\left( 0\right) =-\varepsilon _{0}\left( 0\right) =g\left(
0\right) /\left( 2a\left( 0\right) \right) $ and $\kappa _{0}\left( 0\right)
=0)$ provided that $\mu _{0}$ and $\mu _{1}$ are the standard solutions of
equation (\ref{CharEq}) corresponding to the following initial conditions $%
\mu _{0}\left( 0\right) =0,$ $\mu _{0}^{\prime }\left( 0\right) =2a\left(
0\right) \neq 0$ and $\mu _{1}\left( 0\right) \neq 0,$ $\mu _{1}^{\prime
}\left( 0\right) =0$ (proofs are outlined in Refs.~\cite{Cor-Sot:Lop:Sua:Sus}%
, \cite{Cor-Sot:SusDPO} and \cite{Suaz:Sus}). (Formulas (\ref{A0})--(\ref{F0}%
) correspond to Green's function of generalized harmonic oscillators; see,
for example, \cite{Cor-Sot:Lop:Sua:Sus}, \cite{Cor-Sot:Sua:SusInv}, \cite%
{Dodonov:Man'koFIAN87}, \cite{LanLopSus11}, \cite{Malkin:Man'ko79}, \cite%
{Suaz:Sus}, \cite{Suslov10} and references therein for more details.)

One may refer to the solution (\ref{A0})--(\ref{F0}) as the fundamental
solution of Riccati-type system (\ref{SysA})--(\ref{SysF}). Thus the
transformation property (\ref{MKernel})--(\ref{FKernel}) allows one to find
solution of the initial value problem in terms of the fundamental solution
(for the nonlinear ODE system).

\section{Integration of the Nonautonomous Linear System}

The transformation (\ref{SolIntTransform}) reduces the linear Schr\"{o}%
dinger equation of generalized harmonic oscillators, namely, equation (\ref%
{SolInt1}) with $h=0,$ to the Schr\"{o}dinger equation for a free particle $%
i\chi _{\tau }=\chi _{\xi \xi }$ with a familiar Green function given by%
\begin{equation}
G\left( \xi ,\eta ,\tau -\tau _{0}\right) =\frac{1}{\sqrt{-4\pi i\left( \tau
-\tau _{0}\right) }}\exp \left[ -i\frac{\left( \xi -\eta \right) ^{2}}{%
4\left( \tau -\tau _{0}\right) }\right] ,  \label{IntLin1}
\end{equation}%
where $\xi =\beta \left( t\right) x+\varepsilon \left( t\right) ,$ $\eta
=\beta \left( 0\right) x+\varepsilon \left( 0\right) $ and $\tau =\gamma
\left( t\right) ,$ $\tau _{0}=\gamma \left( 0\right) .$ One can verify
directly that Green's functions of generalized harmonic oscillators \cite%
{Cor-Sot:Lop:Sua:Sus},%
\begin{equation}
G\left( x,y,t\right) =\frac{1}{\sqrt{2\pi i\mu _{0}\left( t\right) }}\exp %
\left[ i\left( \alpha _{0}\left( t\right) x^{2}+\beta _{0}\left( t\right)
xy+\gamma _{0}\left( t\right) y^{2}+\delta _{0}\left( t\right) x+\varepsilon
_{0}\left( t\right) y+\kappa _{0}\left( t\right) \right) \right] ,
\label{IntLin2}
\end{equation}%
are derived from the simplest free particle propagator (\ref{IntLin1}) with
the help of our transformations (\ref{MKernel})--(\ref{FKernel}). In is
worth noting, though, that the transformation (\ref{SolIntTransform})
requires a knowledge of the functions $\mu ,$ $\alpha ,$ $\beta ,$ $\gamma ,$
$\delta ,$ $\varepsilon $ and $\kappa ,$ which allows to determine the
Green's function for the generalized harmonic oscillators directly from (\ref%
{SolInt2}). Thus finding of this transformation is equivalent to integration
of the original linear equation from very beginning. (Lemma~1 extends this
observation to the nonlinear Schr\"{o}dinger equation (\ref{SolInt2})).

Then the superposition principle allows to solve the corresponding Cauchy
initial value problem:%
\begin{equation}
\psi \left( x,t\right) =\int_{-\infty }^{\infty }G\left( x,y,t\right) \psi
\left( y,0\right) \ dy  \label{IntLin3}
\end{equation}%
for suitable initial data $\psi \left( x,0\right) =\varphi \left( x\right) $
(see Refs.~\cite{Cor-Sot:Lop:Sua:Sus}, \cite{Suslov10} and \cite{Suaz:Sus}
for details).

As shown in \cite{Suaz:Sus}, the following asymptotics hold%
\begin{eqnarray}
&&\alpha _{0}\left( t\right) =\frac{1}{4a\left( 0\right) t}-\frac{c\left(
0\right) }{4a\left( 0\right) }-\frac{a^{\prime }\left( 0\right) }{%
8a^{2}\left( 0\right) }+\mathcal{O}\left( t\right) ,  \label{CoeffAsymps} \\
&&\beta _{0}\left( t\right) =-\frac{1}{2a\left( 0\right) t}+\frac{a^{\prime
}\left( 0\right) }{4a^{2}\left( 0\right) }+\mathcal{O}\left( t\right) ,
\notag \\
&&\gamma _{0}\left( t\right) =\frac{1}{4a\left( 0\right) t}+\frac{c\left(
0\right) }{4a\left( 0\right) }-\frac{a^{\prime }\left( 0\right) }{%
8a^{2}\left( 0\right) }+\mathcal{O}\left( t\right) ,  \notag \\
&&\delta _{0}\left( t\right) =\frac{g\left( 0\right) }{2a\left( 0\right) }+%
\mathcal{O}\left( t\right) ,\qquad \varepsilon _{0}\left( t\right) =-\frac{%
g\left( 0\right) }{2a\left( 0\right) }+\mathcal{O}\left( t\right) ,  \notag
\\
&&\kappa _{0}\left( t\right) =\mathcal{O}\left( t\right)  \notag
\end{eqnarray}%
as $t\rightarrow 0$ for sufficiently smooth coefficients. Then%
\begin{eqnarray}
G\left( x,y,t\right) &\sim &\frac{1}{\sqrt{2\pi ia\left( 0\right) t}}\exp %
\left[ i\frac{\left( x-y\right) ^{2}}{4a\left( 0\right) t}\right]
\label{GreenAsymp} \\
&&\times \exp \left[ -i\left( \frac{a^{\prime }\left( 0\right) }{%
8a^{2}\left( 0\right) }\left( x-y\right) ^{2}+\frac{c\left( 0\right) }{%
4a\left( 0\right) }\left( x^{2}-y^{2}\right) -\frac{g\left( 0\right) }{%
2a\left( 0\right) }\left( x-y\right) \right) \right]  \notag
\end{eqnarray}%
as $t\rightarrow 0,$ which corrects a typo in Ref.~\cite{Cor-Sot:Lop:Sua:Sus}%
. (Here, $f\sim g$ as $t\rightarrow 0,$ if $\lim_{t\rightarrow 0}\left(
f/g\right) =$ $1.$)

Another form of solution of the linear problem can be found by an
eigenfunction expansion \cite{LanLopSus11}, \cite{Suslov10}.

Numerous examples of (super) integrable (driven) generalized harmonic
oscillators with a detailed bibliography can be found, for instance, in
recent publications \cite{Cor-Sot:Lop:Sua:Sus}, \cite{Cor-Sot:Sua:Sus}, \cite%
{Cor-Sot:Sua:SusInv}, \cite{Cor-Sot:Sus} and \cite{Cor-Sot:SusDPO}. In
addition, our Lemma~1 shows that these results provide explicit
transformations (\ref{SolIntTransform}) of the corresponding nonlinear
systems into the standard completely integrable forms.

\section{One Soliton Solution}

In the next few sections, we summarize basics of the inverse scattering
technique for the autonomous nonlinear Schr\"{o}dinger equation (\ref%
{SolInt3}) in order to facilitate use of the transformation (\ref%
{SolIntTransform}) for specific nonautonomous and inhomogeneous nonlinear
Schr\"{o}dinger equations (\ref{SolInt2}) from\ various applications. More
details, when needed, can be found in classical works \cite{AblowClark91},
\cite{Ablo:Seg81}, \cite{Fadd:Takh}, \cite{Novikovetal}, \cite{Scott03},
\cite{Zakh:Shab71}, \cite{ZakhShab74}, \cite{ZakhShab79} (see also
references cited in the introduction).

As well-known, equation (\ref{SolInt3}) has a travelling wave solution of
the form%
\begin{equation}
\chi \left( \xi ,\tau \right) =e^{i\left( \xi y+\tau \left(
y^{2}-g_{0}\right) +\phi \right) }F\left( \xi +2\tau y\right)  \label{OneSol}
\end{equation}%
provided%
\begin{equation}
\left( \frac{dF}{dz}\right) ^{2}=C_{0}+g_{0}F^{2}+\frac{1}{2}%
h_{0}F^{4}\qquad \left( C_{0}\text{ is a constant of integration}\right) .
\label{EllipticFuncs}
\end{equation}
Examples include bright and dark solitons, and Jacobi elliptic
transcendental solutions for solitary wave profiles \cite{AblowClark91},
\cite{KudryashovBook10}, \cite{Novikovetal}, \cite{Scott03}, \cite%
{SuazoSuslovSol}. Setting $C_{0}=y=0,$ gives the stationary breather, which
is located about $\xi =0$ and oscillates at a frequency equal to $g_{0}$
\cite{SatYaj74}, \cite{Scott03}.

By (\ref{SolIntTransform}), the nonautonomous Schr\"{o}dinger equation (\ref%
{SolInt2}) under the integrability condition (\ref{SolInt2a}) has the
following solution:%
\begin{eqnarray}
\psi \left( x,t\right) &=&\frac{e^{i\phi }}{\sqrt{\mu }}\exp \left( i\left(
\alpha x^{2}+\beta xy+\gamma \left( y^{2}-g_{0}\right) +\delta x+\varepsilon
y+\kappa \right) \right)  \label{OneSolSol} \\
&&\times F\left( \beta x+2\gamma y+\varepsilon \right) ,  \notag
\end{eqnarray}%
where the elliptic function $F$ satisfies equation (\ref{EllipticFuncs}) and
$\phi ,$ $y,$ $g_{0}$ and $h_{0}$ are real parameters (see also Ref.~\cite%
{SuazoSuslovSol} for a direct derivation of this solution).

\section{Integrability of Nonautonomous Nonlinear Schr\"{o}dinger Equation}

The substitution $\Psi \left( X,T\right) =\sqrt{h_{0}}\chi \left( \sqrt{2}%
X,-2T\right) $ transforms equation (\ref{SolInt3}) into the standard forms%
\begin{equation}
i\Psi _{T}+\Psi _{XX}\pm 2\left\vert \Psi \right\vert ^{2}\Psi =0
\label{StNonLinSchrEq}
\end{equation}%
(focusing and defocusing), which can be obtained as the flatness condition:%
\begin{equation}
U_{T}-V_{X}+UV-VU=0  \label{Lax}
\end{equation}%
for the Lax--(Zakharov--Shabat) pair:%
\begin{eqnarray}
U &=&-i\lambda \sigma _{3}+\Psi \sigma _{+}\mp \Psi ^{\ast }\sigma _{-}
\label{LZSU} \\
&=&\left(
\begin{array}{cc}
-i\lambda & \Psi \\
\mp \Psi ^{\ast } & i\lambda%
\end{array}%
\right)  \notag
\end{eqnarray}%
and%
\begin{eqnarray}
V &=&i\left( -2\lambda ^{2}\pm \left\vert \Psi \right\vert ^{2}\right)
\sigma _{3}+\left( 2\lambda \Psi +i\Psi _{X}\right) \sigma _{+}\pm \left(
-2\lambda \Psi ^{\ast }+i\Psi _{X}^{\ast }\right) \sigma _{-}  \label{LZSV}
\\
&=&\left(
\begin{array}{cc}
i\left( -2\lambda ^{2}\pm \left\vert \Psi \right\vert ^{2}\right) & 2\lambda
\Psi +i\Psi _{X} \\
\mp 2\lambda \Psi ^{\ast }\pm i\Psi _{X}^{\ast } & i\left( 2\lambda ^{2}\mp
\left\vert \Psi \right\vert ^{2}\right)%
\end{array}%
\right)  \notag
\end{eqnarray}%
(we use the asterisk for complex conjugation). Here, $\lambda $ is a
constant, $\sigma _{\pm }=\left( \sigma _{1}\pm i\sigma _{2}\right) /2$ and $%
\sigma _{1},$ $\sigma _{2},$ $\sigma _{3}$ are the Pauli matrices:%
\begin{equation}
\sigma _{1}=\left(
\begin{array}{cc}
0 & 1 \\
1 & 0%
\end{array}%
\right) ,\qquad \sigma _{2}=\left(
\begin{array}{cc}
0 & -i \\
i & 0%
\end{array}%
\right) ,\qquad \sigma _{3}=\left(
\begin{array}{cc}
1 & 0 \\
0 & -1%
\end{array}%
\right) .  \label{PauliSigmas}
\end{equation}%
Since the Lax pair guarantees complete integrability and can alone derive
all its associated properties, our transformation (\ref{SolIntTransform})
trivially explains the integrability features of the nonautonomous nonlinear
Schr\"{o}dinger equation (\ref{SolInt1}), including $N$-soliton solutions,
infinite conservative properties, etc., (see Refs.~\cite{Kundu09}, \cite%
{Novikovetal} and \cite{Scott03} for more details).

\section{Inverse Scattering Method}

Solution of the Cauchy initial value problem through the inverse scattering
method is discussed in \cite{AblowPrinTrub04}, \cite{Ablo:Seg81}, \cite%
{KudryashovBook10}, \cite{Novikovetal}, \cite{Scott03}, \cite{Zakh:Shab71},
\cite{ZakhShab74} and \cite{ZakhShab79}. In the focusing case,%
\begin{equation}
i\Psi _{T}+\Psi _{XX}+2\left\vert \Psi \right\vert ^{2}\Psi =0,
\label{NLSEqFocusing}
\end{equation}%
the Zakharov--Shabat system contains four equations for an auxiliary
two-component wave function $\boldsymbol{\Phi }=\left( \varphi ,\upsilon
\right) ^{T}:$%
\begin{equation}
\boldsymbol{\Phi }_{X}=U\boldsymbol{\Phi },\qquad \boldsymbol{\Phi }_{T}=V%
\boldsymbol{\Phi },  \label{ZShSys}
\end{equation}%
namely,%
\begin{eqnarray}
\varphi _{X} &=&-i\lambda \varphi +\Psi \upsilon ,  \label{LaxPairI} \\
\upsilon _{X} &=&-\Psi ^{\ast }\varphi +i\lambda \upsilon  \notag
\end{eqnarray}%
and%
\begin{eqnarray}
\varphi _{T} &=&i\left( -2\lambda ^{2}+\left\vert \Psi \right\vert
^{2}\right) \varphi +\left( 2\lambda \Psi +i\Psi _{X}\right) \upsilon ,
\label{LaxPairII} \\
\upsilon _{T} &=&\left( -2\lambda \Psi ^{\ast }+i\Psi _{X}^{\ast }\right)
\varphi +i\left( 2\lambda ^{2}-\left\vert \Psi \right\vert ^{2}\right)
\upsilon .  \notag
\end{eqnarray}%
Assuming that $\Psi \left( X,T\right) \rightarrow 0$ $\left( \text{and }\Psi
_{X}\left( X,T\right) \rightarrow 0\right) $ and as $X\rightarrow \pm \infty
$ implies%
\begin{equation}
\varphi _{T}\rightarrow -2i\lambda ^{2}\varphi ,\qquad \upsilon
_{T}\rightarrow 2i\lambda ^{2}\upsilon \qquad \left( X\rightarrow \pm \infty
\right)  \label{LaxPairAsympt}
\end{equation}%
so the scattering data for the problem%
\begin{eqnarray}
&&L\left(
\begin{array}{c}
\varphi \\
\upsilon%
\end{array}%
\right) =\lambda \left(
\begin{array}{c}
\varphi \\
\upsilon%
\end{array}%
\right) ,  \label{EVProbL} \\
&&L=i\sigma _{3}\frac{\partial }{\partial X}-i\Psi \sigma _{+}-i\Psi ^{\ast
}\sigma _{-}=i\left(
\begin{array}{cc}
\partial _{X} & -\Psi \\
-\Psi ^{\ast } & -\partial _{X}%
\end{array}%
\right)  \label{LaxL}
\end{eqnarray}%
evolve with time as%
\begin{equation}
b\left( \lambda ,T\right) =b\left( \lambda ,T_{0}\right) e^{4i\lambda
^{2}\left( T-T_{0}\right) },\qquad r_{n}\left( T\right) =r_{n}\left(
T_{0}\right) e^{4i\lambda \left( T-T_{0}\right) }.  \label{AsymptBR}
\end{equation}

Then Cauchy initial value problem for the nonlinear Schr\"{o}dinger equation
(\ref{StNonLinSchrEq}) can be solved as follows \cite{KudryashovBook10},
\cite{Scott03}, \cite{Zakh:Shab71}, \cite{ZakhShab74}, \cite{ZakhShab79}:
\begin{equation}
\Psi \left( X,T\right) =-2K\left( X,X,T\right) ,  \label{SolCIVP}
\end{equation}%
where $K\left( X,Y,T\right) $ satisfies the linear integral equation%
\begin{eqnarray}
K\left( X,Y,T\right) &=&B^{\ast }\left( X+Y,T\right)  \label{GLMEq} \\
&&-\int_{X}^{\infty }\int_{X}^{\infty }K\left( X,Z,T\right) B\left(
Z+W,T\right) B^{\ast }\left( Y+W,T\right) \ dZdW  \notag
\end{eqnarray}%
and $B\left( X,T\right) $ can be obtained in terms of the scattering data:%
\begin{equation}
B\left( X,T\right) =-i\sum_{n=1}^{N}r_{n}\left( T_{0}\right) e^{i\left(
\lambda _{n}X+4\lambda _{n}^{2}\left( T-T_{0}\right) \right) }+\frac{1}{2\pi
}\int_{-\infty }^{\infty }b\left( \lambda ,T_{0}\right) e^{i\left( \lambda
X+4\lambda ^{2}\left( T-T_{0}\right) \right) }\ d\lambda  \label{BKernelISM}
\end{equation}%
(see Refs.~\cite{AblowClark91}, \cite{AblowPrinTrub04}, \cite{Ablo:Seg81},
\cite{KudryashovBook10}, \cite{Novikovetal}, \cite{SatYaj74}, \cite{Scott03}%
, \cite{Tao09}, \cite{Zakh:Shab71}, \cite{ZakhShab74}, \cite{ZakhShab79} for
more details).

As a result, the following combination of gauge, scaling and coordinate
transformations \cite{Kundu09}:%
\begin{equation}
\psi \left( x,t\right) =\frac{1}{\sqrt{h_{0}\mu \left( t\right) }}e^{i\left(
\alpha \left( t\right) x^{2}+\delta \left( t\right) x+\kappa \left( t\right)
\right) }\ \Psi \left( \frac{1}{\sqrt{2}}\left( \beta \left( t\right)
x+\varepsilon \left( t\right) \right) ,-\frac{1}{2}\gamma \left( t\right)
\right) ,  \label{CIVPSol}
\end{equation}%
written here explicitly in terms of solution of the Riccati-type system (\ref%
{SysA})--(\ref{SysF}) from section~3, allows to solve Cauchy initial value
problem for the nonautonomous nonlinear Schr\"{o}dinger equation (\ref%
{SolInt2}) with the help of the standard inverse scattering technique. The
following choice of parameters: $\alpha \left( 0\right) =\gamma \left(
0\right) =\delta \left( 0\right) =\kappa \left( 0\right) =\varepsilon \left(
0\right) =0,$ $\beta \left( 0\right) =\sqrt{2},$ $h_{0}\mu \left( 0\right)
=1 $ preserves the initial data: $\psi \left( x,0\right) =\Psi \left( \xi
,0\right) $ and simplifies the solution (\ref{MKernel})--(\ref{FKernel}).

\section{Two Soliton Solution}

Two well-known solutions of (\ref{NLSEqFocusing}) are given by \cite%
{SatYaj74}, \cite{Scott03}:
\begin{equation}
\Psi _{1}\left( X,T\right) =\frac{e^{iT}}{\cosh X}  \label{OneSoliton}
\end{equation}%
and%
\begin{equation}
\Psi _{2}\left( X,T\right) =4e^{iT}\frac{\cosh 3X+3e^{8iT}\cosh X}{\cosh
4X+4\cosh 2X+3\cos 8T}.  \label{TwoSolitons}
\end{equation}%
Use of the transformation (\ref{CIVPSol}), results in one and two soliton
solutions for the nonautonomous nonlinear Schr\"{o}dinger equation (\ref%
{SolInt2}), respectively. (See \cite{SatYaj74}, \cite{Scott03} and \cite%
{Zakh:Shab71}, for more details; $N$-soliton solutions are also discussed in
\cite{AblowClark91}, \cite{Mateev:SalleBook} and \cite{Novikovetal}.)

\section{Transformation of the Lax Pair and Zakharov--Shabat System}

If needed, an equivalent (nonisospectral) Lax pair for the nonautonomous Schr%
\"{o}dinger equation (\ref{SolInt2}), which is discussed in Refs.~\cite%
{Balakrish85}, \cite{ChenHH:LiuCS76}, \cite{ChenHH:LiuCS78}, \cite{Clark88},
\cite{Serkinetal07}, \cite{Serkinetal10} for important special cases (see
also \cite{AlKha10} and \cite{Bruga:Sci10}), can be derived, in general,
from (\ref{LZSU})--(\ref{LZSV}) by inverting our transformation (\ref%
{CIVPSol}) (see \cite{AblowPrinTrub04}, \cite{Kundu09} for more details). In
this paper, the required integrability condition (\ref{SolInt2a}) (found for
the soliton-like solution in Ref.~\cite{SuazoSuslovSol}) has been already
incorporated into this transformation. Computational details are left to the
reader.

\section{Examples}

A few simple examples are in order. (More examples can be found in \cite%
{Cor-Sot:Sua:SusInv}, \cite{SuazoSuslovSol}; see also references therein.)

\subsection{Example~1}

As noticed by Clarkson \cite{Clark88} (see also \cite{AblowClark91} and \cite%
{Musette99}), the equation%
\begin{equation}
i\psi _{t}+\psi _{xx}=\frac{\omega ^{2}}{4}x^{2}\psi +2\left\vert \psi
\right\vert ^{2}\psi ,\qquad \omega \neq 0  \label{clark1}
\end{equation}%
does not pass the Painlev\'{e} test and, therefore, is not integrable. The
corresponding characteristic equation, $\mu ^{\prime \prime }+\omega ^{2}\mu
=0,$ has two standard solutions:%
\begin{equation}
\mu _{0}=\frac{2}{\omega }\sin \omega t,\qquad \mu _{1}=\cos \omega t.
\label{clark2}
\end{equation}%
By our Lemma~1, a modified equation for harmonic solitons \cite{GagWint93},
\cite{SuazoSuslovSol}:%
\begin{equation}
i\psi _{t}+\psi _{xx}=\frac{\omega ^{2}}{4}x^{2}\psi +\frac{h_{0}\omega \mu
\left( 0\right) \beta ^{2}\left( 0\right) }{4\alpha \left( 0\right) \sin
\omega t+\omega \cos \omega t}\left\vert \psi \right\vert ^{2}\psi
\label{clark3}
\end{equation}%
($h_{0},$ $\alpha \left( 0\right) ,$ $\beta \left( 0\right) \neq 0$ and $\mu
\left( 0\right) $ are real constants) can be transformed into the standard
form and hence is integrable. Here, $\mu \left( t\right) =\mu \left(
0\right) \left[ 4\alpha \left( 0\right) \sin \omega t+\omega \cos \omega t%
\right] /\omega $ and general solution of the corresponding Riccati-type
system is given by%
\begin{eqnarray}
\alpha \left( t\right)  &=&\frac{\omega }{4}\frac{4\alpha \left( 0\right)
\cos \omega t-\omega \sin \omega t}{4\alpha \left( 0\right) \sin \omega
t+\omega \cos \omega t}, \\
\beta \left( t\right)  &=&\frac{\omega \beta \left( 0\right) }{4\alpha
\left( 0\right) \sin \omega t+\omega \cos \omega t}, \\
\gamma \left( t\right)  &=&\gamma \left( 0\right) -\frac{\beta ^{2}\left(
0\right) \sin \omega t}{4\alpha \left( 0\right) \sin \omega t+\omega \cos
\omega t}, \\
\delta \left( t\right)  &=&\frac{\omega \delta \left( 0\right) }{4\alpha
\left( 0\right) \sin \omega t+\omega \cos \omega t}, \\
\varepsilon \left( t\right)  &=&\varepsilon \left( 0\right) -\frac{2\beta
\left( 0\right) \delta \left( 0\right) \sin \omega t}{4\alpha \left(
0\right) \sin \omega t+\omega \cos \omega t}, \\
\kappa \left( t\right)  &=&\kappa \left( 0\right) -\frac{\delta ^{2}\left(
0\right) \sin \omega t}{4\alpha \left( 0\right) \sin \omega t+\omega \cos
\omega t}.
\end{eqnarray}%
Letting $\mu \left( 0\right) =\beta \left( 0\right) =1$ and $\alpha \left(
0\right) =\gamma \left( 0\right) =\delta \left( 0\right) =\varepsilon \left(
0\right) =\kappa \left( 0\right) =0,$ we arrive at the simple substitution:%
\begin{equation}
\psi \left( x,t\right) =\frac{e^{-\left( \omega /4\right) x^{2}\tan \omega t}%
}{\sqrt{\cos \omega t}}\ \chi \left( \frac{x}{\cos \omega t},-\frac{\tan
\omega t}{\omega }\right) ,  \label{ClarkWint}
\end{equation}%
which transforms (\ref{clark3}) into (\ref{SolInt3}). (Replace $\omega
=i\varkappa $ for the original equation in Ref.~\cite{GagWint93}.)

In a similar fashion, the following equation%
\begin{equation}
i\psi _{t}+\psi _{xx}+\left( k^{2}x^{2}-ik\right) \psi =\frac{2kh_{0}\mu
\left( 0\right) \beta ^{2}\left( 0\right) }{\left( k+2\alpha \left( 0\right)
\right) e^{4kt}+k-2\alpha \left( 0\right) }\left\vert \psi \right\vert
^{2}\psi  \label{clark4}
\end{equation}%
($k\neq 0,$ $\alpha \left( 0\right) ,$ $\beta \left( 0\right) \neq 0$ and $%
\mu \left( 0\right) $ are real constants) is integrable. Indeed, the
characteristic equation and the standard solutions are given by%
\begin{equation}
\mu ^{\prime \prime }+4k\mu ^{\prime }=0;\qquad \mu _{0}=\frac{1}{2k}\left(
1-e^{-4kt}\right) ,\quad \mu _{1}=1  \label{clark5}
\end{equation}%
and%
\begin{equation}
\mu =\frac{\mu \left( 0\right) }{2k}\left[ k+2\alpha \left( 0\right) +\left(
k-2\alpha \left( 0\right) \right) e^{-4kt}\right] ,\qquad \lambda =e^{-2kt}.
\label{clark6}
\end{equation}%
The simplest case occurs when $k+2\alpha \left( 0\right) =0$ \cite{Clark88}.
Further details are left to the reader.

\subsection{Example~2}

A soliton moving with acceleration in linearly inhomogeneous plasma was
discovered in Refs.~\cite{ChenHH:LiuCS76} and \cite{ChenHH:LiuCS78} (see
also \cite{Balakrish85}, \cite{TappZab71}). For a modified equation with the
integrability condition (\ref{SolInt2a}),%
\begin{equation}
i\frac{\partial \psi }{\partial t}+\frac{\partial ^{2}\psi }{\partial x^{2}}%
+2kx\psi =\frac{h_{0}\mu _{0}\beta _{0}^{2}}{1+4\alpha _{0}t}\left\vert \psi
\right\vert ^{2}\psi ,  \label{ihp1}
\end{equation}%
where $k,$ $h_{0},$ $\alpha _{0},$ $\beta _{0}$ and $\mu _{0}$ are
constants, we get $\mu \left( t\right) =\mu _{0}\left( 1+4\alpha
_{0}t\right) $ and%
\begin{eqnarray}
&&\alpha \left( t\right) =\frac{\alpha _{0}}{1+4\alpha _{0}t},\qquad \qquad
\quad \beta \left( t\right) =\frac{\beta _{0}}{1+4\alpha _{0}t},
\label{ihp2} \\
&&\gamma \left( t\right) =\gamma _{0}-\frac{\beta _{0}^{2}t}{1+4\alpha _{0}t}%
,\quad \qquad \delta \left( t\right) =kt+\frac{\delta _{0}+kt}{1+4\alpha
_{0}t},  \label{ihp3} \\
&&\varepsilon \left( t\right) =\varepsilon _{0}-\frac{2\beta _{0}t\left(
\delta _{0}+kt\right) }{1+4\alpha _{0}t},\quad \kappa \left( t\right)
=\kappa _{0}-\frac{k^{2}t^{3}}{3}-\frac{t\left( \delta _{0}+kt\right) ^{2}}{%
1+4\alpha _{0}t}.  \label{ihp4}
\end{eqnarray}%
The classical case \cite{ChenHH:LiuCS76}, \cite{ChenHH:LiuCS78} corresponds
to $\alpha _{0}=0$ and $h_{0}\mu _{0}\beta _{0}^{2}=-2$ (with $k\rightarrow
-k).$ The simplest transformation:%
\begin{equation}
\psi \left( x,t\right) =e^{i\left( 2ktx-4k^{2}t^{3}/3\right) }\ \chi \left(
x-2kt^{2},-t\right) ,  \label{Tappert}
\end{equation}%
when $\alpha =\alpha _{0}=\gamma _{0}=\delta _{0}=\varepsilon _{0}=\kappa
_{0}=\mu _{0}=0$ and $\beta =1,$ is due to Tappert \cite{ChenHH:LiuCS76}.

\subsection{Example~3}

The nonlinear Schr\"{o}dinger equation \cite{SuazoSuslovSol}:%
\begin{equation}
i\frac{\partial \psi }{\partial t}+\frac{\partial ^{2}\psi }{\partial x^{2}}=%
\frac{g_{0}\beta _{0}^{2}}{\left( 1+4\alpha _{0}t\right) ^{2}}z\psi +\frac{%
h_{0}\mu _{0}\beta _{0}^{2}}{1+4\alpha _{0}t}\left\vert \psi \right\vert
^{2}\psi ,  \label{ihp4a}
\end{equation}%
where%
\begin{equation}
z=\frac{\beta _{0}x+2\left( \gamma _{0}-\left( \beta _{0}^{2}-4\alpha
_{0}\gamma _{0}\right) t\right) y}{1+4\alpha _{0}t},  \label{ihp5}
\end{equation}%
with the help of the gauge transformation:%
\begin{equation}
\psi =e^{-if\left( t\right) }\chi \left( x,t\right) ,\qquad \frac{df}{dt}%
=2g_{0}\beta _{0}^{2}y\frac{\gamma _{0}-\left( \beta _{0}^{2}-4\alpha
_{0}\gamma _{0}\right) t}{\left( 1+4\alpha _{0}t\right) ^{3}},  \label{ihp6}
\end{equation}%
can be transformed into a similar form
\begin{equation}
i\chi _{t}+\chi _{xx}-\frac{g_{0}\beta _{0}^{3}x}{\left( 1+4\alpha
_{0}t\right) ^{3}}\chi =\frac{h_{0}\mu _{0}\beta _{0}^{2}}{1+4\alpha _{0}t}%
\left\vert \chi \right\vert ^{2}\chi .  \label{ihp7a}
\end{equation}%
Looking for a traveling wave, we indicate the following soliton-like
solution \cite{SuazoSuslovSol}, \cite{Tajiri83}:%
\begin{equation}
\chi \left( x,t\right) =\frac{e^{iS\left( x,t\right) }}{\sqrt{\left\vert \mu
_{0}\right\vert \left( 1+4\alpha _{0}t\right) }}g_{0}^{1/3}\sqrt{\frac{2}{%
h_{0}}}A_{k_{0}}\left( g_{0}^{1/3}z\right) ,  \label{ihp8}
\end{equation}%
where%
\begin{eqnarray}
S\left( x,t\right) &=&\frac{\alpha _{0}x^{2}+\beta _{0}xy+\left( \gamma
_{0}-\left( \beta _{0}^{2}-4\alpha _{0}\gamma _{0}\right) t\right) y^{2}}{%
1+4\alpha _{0}t}  \label{ihp9} \\
&&+g_{0}\beta _{0}^{2}t\frac{2\gamma _{0}-\left( \beta _{0}^{2}-8\alpha
_{0}\gamma _{0}\right) t}{1+4\alpha _{0}t}y  \notag
\end{eqnarray}%
($\alpha _{0},$ $\beta _{0},$ $\gamma _{0},$ $\delta _{0},$ $\varepsilon
_{0},$ $\kappa _{0},$ $\mu _{0},$ $g_{0},$ $h_{0},$ $y$ are real constants)
and the soliton profile is defined, as a solution of the second Painlev\'{e}
equation \cite{Tajiri83}, in terms of the nonlinear Airy function $%
A_{k_{0}}\left( \zeta \right) $ with asymptotics given by%
\begin{equation}
A_{k_{0}}\left( \zeta \right) =\left\{
\begin{array}{c}
k_{0}\text{Ai\/}\left( \zeta \right) ,\qquad \zeta \rightarrow +\infty
\bigskip \\
r\left\vert \zeta \right\vert ^{-1/4}\sin \left( s\left( \zeta \right)
-\theta _{0}\right) +\text{o}\left( \left\vert \zeta \right\vert
^{-1/4}\right) ,\quad \zeta \rightarrow -\infty \bigskip .%
\end{array}%
\right.  \label{AsPII}
\end{equation}%
Here, Ai\/$\left( \zeta \right) $ is the Ai\/ry function, $-1<k_{0}<1$
provided $k_{0}\neq 0,$ $r^{2}=-\pi ^{-1}\ln \left( 1-k_{0}^{2}\right) ,$%
\begin{equation}
s\left( \zeta \right) =\frac{2}{3}\left\vert \zeta \right\vert ^{3/2}-\frac{3%
}{4}r^{2}\ln \left\vert \zeta \right\vert  \label{AsPIIs}
\end{equation}%
and%
\begin{equation}
\theta _{0}=\frac{3}{2}r^{2}\ln 2+\arg \Gamma \left( 1-\frac{i}{2}%
r^{2}\right) +\frac{\pi }{4}\left( 1-2\text{sign\/}\left( k_{0}\right)
\right) .  \label{AsPIIph}
\end{equation}%
These asymptotics were found in \cite{AbloSeg77}, \cite{SegurAb81} and had
been proven rigorously in \cite{DeiftZhou93}, \cite{DeiftZhou95} (see \cite%
{Ablo:Seg81}, \cite{Bassometal98}, \cite{Clark10}, \cite{ClarkMcLeod88},
\cite{Conte99}, \cite{ConteMusette09}, \cite{Takei02} and references therein
for study of this nonlinear Airy function).

It is worth noting that, in a contrast to the previous case, this $A$%
-soliton moves with a constant velocity when $\alpha _{0}=0$ (notice that
the external field has essentially changed the soliton shape; see Ref.~\cite%
{SuazoSuslovSol} for more details).

\section{Conclusion}

We have shown how to transform the nonautonomous and inhomogeneous nonlinear
Schr\"{o}dinger equation (\ref{SolInt2})--(\ref{SolInt2a}) into the standard
autonomous form (\ref{SolInt3}) that is completely integrable by the inverse
scattering approach. This transformation is explicitly written in terms of
Green's function of the corresponding linear problem (generalized harmonic
oscillators); see Lemma~1 and equations (\ref{MKernel})--(\ref{FKernel}) and
(\ref{A0})--(\ref{F0}). Combination of these advances allows one to solve
the Cauchy initial value problem for the generic nonautonomous integrable
quantum system under consideration.\ Simple examples are considered and
further examples can be found elsewhere (see, for instance, \cite%
{Cor-Sot:Sua:SusInv}, \cite{SuazoSuslovSol} and references therein).

\noindent \textbf{Acknowledgments.\/} The author thanks George E. Andrews,
Carlos Castillo-Ch\'{a}vez, Robert Conte, Vladimir I. Man'ko and Svetlana
Roudenko for support, valuable discussions and encouragement. I am grateful
to Nathan Lanfear and Erwin Suazo for help. This work is supported in part
by the National Science Foundation--Enhancing the Mathematical Sciences
Workforce in the 21st Century (EMSW21), award \# 0838705; the Alfred P.
Sloan Foundation--Sloan National Pipeline Program in the Mathematical and
Statistical Sciences, award \# LTR 05/19/09; and the National Security
Agency--Mathematical \& Theoretical Biology Institute---Research program for
Undergraduates; award \# H98230-09-1-0104.


\begin{thebibliography}{999}
\bibitem{Ablowitzetal73} M.~J.~Ablowitz, D.~J.~Kaup, A.~C.~Newell and
H.~Segur, \emph{Nonlinear-evolution equations of physical significance\/},
Phys. Rev. Lett. \textbf{31} (1973)~\#2, 125--127.

\bibitem{AblowClark91} M.~Ablowitz and P.~A.~Clarkson, \textsl{Solitons,
Nonlinear Evolution Equations and Inverse Scattering\/}, Cambridge Univ.
Press, Cambridge, 1991.

\bibitem{AblowPrinTrub04} M.~Ablowitz, B.~Prinari and A.~D.~Trubatch,
\textsl{Discrete and Continuous Schr\"{o}dinger Systems\/},
Cambridge Univ. Press, Cambridge, 2004.

\bibitem{AbloSeg77} M.~Ablowitz and H.~Segur, \emph{Exact linearization of a
Painlev\'{e} transcendent\/}, Phys. Rev. Lett. \textbf{38} (1977)~\#20,
1103--1106.

\bibitem{Ablo:Seg81} M.~Ablowitz and H.~Segur, \textsl{Solitons and the
Inverse Scattering Transform\/}, SIAM, Philadelphia, 1981.

\bibitem{Agrawal} G.~P.~Agrawal, \textsl{Nonlinear Fiber Optics\/}, Forth
Edition, Academic Press, New York, 2007.

\bibitem{AlKha10} U.~Al~Khawaja, \emph{A comparative analysis of Painlev\'{e}%
, Lax pair, and similarity transformation methods in obtaining the
integrability conditions of nonlinear Schr\"{o}dinger equations\/}, J. Math.
Phys. \textbf{51} (2010), 053506 (11 pages).

\bibitem{Atre:Pani:Aga06} R.~Atre, P.~K.~Panigrahi and G.~S.~Agarwal, \emph{%
Class of solitary wave solutions of the one-dimensional Gross--Pitaevskii
equation\/},~Phys. Rev. E \textbf{73} (2006), 056611.

\bibitem{Balakrish85} R.~Balakrishnan, \emph{Soliton propagation in
nonunivorm media\/},~Phys. Rev. A \textbf{32} (1985)~\#2, 1144--1149.

\bibitem{Bassometal98} A.~P.~Bassom, P.~A.~Clarkson, C.~K.~Law and
J.~B.~McLeod, \emph{Application of uniform asymptotics to the second Painlev%
\'{e} transcendent\/}, Arch. Rat. Mech. Anal. \textbf{103} (1998), 241--271.

\bibitem{Berry85} M.~V.~Berry, \emph{Classical adiabatic angles and quantum
adiabatic phase\/}, J. Phys.~A: Math. Gen. \textbf{18} (1985) \# 1, 15--27.

\bibitem{BoitiGuiz82} M.~Boiti and T.~Guizhang, \emph{B\"{a}cklund
transformations via gauge transformations\/}, Nuovo Cimento B \textbf{71}
(1982)~\#~2, 253--264.

\bibitem{BoitiPem80} M.~Boiti and F.~Pempinelli, \emph{Nonlinear Schr\"{o}%
dinger equation, B\"{a}cklund transformations and Painlev\'{e}
transcendents\/}, Nuovo Cimento B \textbf{59} (1980)~\# 1, 40--58.

\bibitem{Bol-BPerezVersKonotop08} J.~Bolmonte-Beitia, V.~M.~P\'{e}rez-Garc%
\'{\i}a, V.~Vekslerchik and V.~V.~Konotop, \emph{Localized nonlinear waves
in systems with time- and space-modulated nonlinearities\/}, Phys. Rev.
Lett. \textbf{100} (2008), 164102 (4~pages).

\bibitem{BongsSengs04} K.~Bongs and K.~Sengstock, \emph{Physics with
coherent matter waves\/}, Rep. Prog. Phys. \textbf{67} (2004), 907--963.

\bibitem{Bruga:Sci10} T.~Brugarino and M.~Sciacca, \emph{Integrability of an
inhomogeneous nonlinear Schr\"{o}dinger equation in Bose--Einstein
condensates and fiber optics\/}, J. Math. Phys. \textbf{51} (2010), 093503
(18 pages).

\bibitem{ChenHH74} H.-H.~Chen, \emph{General deriviation of B\"{a}cklund
transformations from inverse scattering problems\/}, Phys. Rev. Lett.
\textbf{33} (1974)~\#15, 925--928.

\bibitem{ChenHH:LiuCS76} H.-H.~Chen and Ch.-Sh.~Liu, \emph{Solitons in
nonuniform media\/}, Phys. Rev. Lett. \textbf{37} (1976)~\#11, 693--697.

\bibitem{ChenHH:LiuCS78} H.-H.~Chen and Ch.-Sh.~Liu, \emph{Nonlinear wave
and soliton propagation in media with arbitrary inhomogeneities\/}, Phys.
Fluids \textbf{21} (1978)~\#3, 377--380.

\bibitem{ChenYi05} Sh.~Chen and L.~Yi, \emph{Chirped self-similar solitons
of a generalized Schr\"{o}dinger equation model}\textit{\/}, Phys. Rev. E
\textbf{61 }(2005), 016606 (4 pages).

\bibitem{Clark88} P.~A.~Clarkson, \emph{Painlev\'{e} analysis for the
damped, driven nonlinear Schr\"{o}dinger equation\/},~Proc. Roy. Soc.~Edin.,
\textbf{109A} (1988), 109--126.

\bibitem{Clark10} P.~A.~Clarkson, \emph{Painlev\'{e} transcendents\/}, in:
\textsl{NIST Handbook of Mathematical Functions\/}, (F.~W.~J.~Olwer,
D.~M.~Lozier \textit{et al}, Eds.), Cambridge Univ. Press, 2010; see also:
http://dlmf.nist.gov/32.

\bibitem{ClarkMcLeod88} P.~A.~Clarkson and J.~B.~McLeod, \emph{A connection
formula for the second Painlev\'{e} transcendent\/}, Arch. Rat. Mech. Anal.
\textbf{103} (1988), 97--138.

\bibitem{Conte89} R.~Conte, \emph{Invariant Painlev\'{e} analysis of partial
differential equations\/},~Phys. Lett.~A \textbf{140} (1989)~\#7--8,
383--390.

\bibitem{Conte99} R.~Conte, \emph{The Painlev\'{e} approach to nonlinear
ordinary differential equations\/},~in: \textsl{The Painlev\'{e} Property,
One Century Later\/}, CRM Series in Mathematical Physics (R.~Conte, Ed.),
Springer--Verlag, New York, 1999. pp.~77--180.

\bibitem{ConteFordyPick93} R.~Conte, A.~P.~Fordy and A.~Pickering, \emph{A
perturbative Painlev\'{e} approach to nonlinear differential equations\/}%
,~Physica~D \textbf{69} (1993), 33--58.

\bibitem{ConteMusette09} R.~Conte and M.~Musette, \emph{Elliptic general
analytic solutions\/},~Studies in Applied Mathematics \textbf{123}
(2009)~\#1, 63--81.

\bibitem{Cor-Sot:Lop:Sua:Sus} R.~Cordero-Soto, R.~M.~Lopez, E.~Suazo and
S.~K.~Suslov, \emph{Propagator of a charged particle with a spin in uniform
magnetic and perpendicular electric fields\/}, Lett.~Math.~Phys. \textbf{84}
(2008)~\#2--3, 159--178.

\bibitem{Cor-Sot:Sua:Sus} R.~Cordero-Soto, E.~Suazo and S.~K.~Suslov, \emph{%
Models of damped oscillators in quantum mechanics\/}, Journal of Physical
Mathematics, \textbf{1} (2009), S090603 (16 pages).

\bibitem{Cor-Sot:Sua:SusInv} R.~Cordero-Soto, E.~Suazo and S.~K.~Suslov,
\emph{Quantum integrals of motion for variable quadratic Hamiltonians\/},
Ann. Phys. \textbf{325} (2010)~\#9, 1884--1912; see also arXiv:0912.4900v9
[math-ph] 19 Mar 2010.

\bibitem{Cor-Sot:Sus} R.~Cordero-Soto and S.~K.~Suslov, \emph{Time reversal
for modified oscillators\/}, Theoretical and Mathematical Physics \textbf{162%
} (2010)~\#3, 286--316; see also arXiv:0808.3149v9 [math-ph] 8~Mar 2009.

\bibitem{Cor-Sot:SusDPO} R.~Cordero-Soto and S.~K.~Suslov, \emph{The
degenerate parametric oscillator and Ince's equation\/}, J.~Phys. A: Math.
Theor. \textbf{44} (2011)~\#1, 015101 (9 pages); see also\emph{\ }%
arXiv:1006.3362v3 [math-ph] 2 Jul 2010.

\bibitem{Cosgrove93I} C.~M.~Cosgrove, \emph{Painlev\'{e} classification of
all semilinear partial differential equations of the second order.
I\/~Hyperbolic equations in two independent variables},~Studies in Applied
Mathematics \textbf{89 }(1993)~\# 1, 1--61.

\bibitem{Cosgrove93II} C.~M.~Cosgrove, \emph{Painlev\'{e} classification of
all semilinear partial differential equations of the second order.
II\/~Parabolic and higher dimentional equations},~Studies in Applied
Mathematics \textbf{89 }(1993)~\# 2, 95--151.

\bibitem{Dal:Giorg:Pitaevski:Str99} F.~Dalfovo, S.~Giorgini,
L.~P.~Pitaevskii and S.~Stringari, \emph{Theory of Bose--Einstein
condensation in trapped gases\/}, Rev. Mod. Phys. \textbf{71} (1999),
463--512.

\bibitem{DegasRes} A.~Desgasperis, \emph{Resource letter Sol-1: Solitons\/},
Am.~J. Phys. \textbf{66} (1998)~6, 486--497.

\bibitem{Degas10} A.~Desgasperis, \emph{Integrable models in nonlinear
optics and soliton solutions\/}, J. Phys.~A: Math. Theor. \textbf{43}
(2010), 434001 (18 pages).

\bibitem{DeiftZhou93} P.~Deift and X.~Zhou, \emph{A steepest descent method
for Riemann--Hilbert problems\/}, Ann.~Math. \textbf{137} (1993), 295--368.

\bibitem{DeiftZhou95} P.~Deift and X.~Zhou, \emph{Asymptotics for Painlev%
\'{e}~II equation\/}, Comm.~Pure Appl.~Math. \textbf{48} (1995), 227--337.

\bibitem{Dod:Mal:Man75} V.~V.~Dodonov, I.~A.~Malkin and V.~I.~Man'ko, \emph{%
Integrals of motion, Green functions, and coherent states of dynamical
systems\/}, Int.~J.~Theor.~Phys. \textbf{14} (1975)~\#~1, 37--54.

\bibitem{Dodonov:Man'koFIAN87} V.~V.~Dodonov and V.~I.~Man'ko, \emph{%
Invariants and correlated states of nonstationary quantum systems}, in:
\textsl{Invariants and the Evolution of Nonstationary Quantum Systems\/},
Proceedings of Lebedev Physics Institute, vol. 183, pp. 71-181, Nauka,
Moscow, 1987 [in Russian]; English translation published by Nova Science,
Commack, New York, 1989, pp. 103-261.

\bibitem{Eba:Khal} A.~Ebaid and S.~M.~Khaled, \emph{New types of exact
solutions for nonlinear Schr\"{o}dinger equation with cubic nonlinearity\/},
Journal of Computational and Applied Mathematics \textbf{235} (2011)~\#8,
1984--1992.

\bibitem{Fadd:Takh} L.~D.~Faddeev and L.~A.~Takhtajan, \textsl{Hamiltonian
Methods in the Theory of Solitons\/}, Springer--Verlag, Berlin, New York,
1987.

\bibitem{Gardneretal67} C.~S.~Gardner, J.~M.~Green, M.~D.~Kruskai and
R.~M.~Miura, \emph{Method for solving the Korteweg--deVries equation\/},
Phys. Rev. Lett. \textbf{19} (1967)~\#19, 1095--1097.

\bibitem{GagWint93} L.~Gagnon and P.~Winternitz, \emph{Symmetry classes of
variable coefficient nonlinear Schr\"{o}dinger equations\/}, J. Phys.~A:
Math. Gen. \textbf{26} (1993), 7061--7076.

\bibitem{Hannay85} J.~H.~Hannay, \emph{Angle variable holonomy in adiabatic
excursion of an integrable Hamiltonian\/}, J. Phys.~A: Math. Gen \textbf{18}
(1985) \# 2, 221--230.

\bibitem{Hasegawa} A.~Hasegawa, \textsl{Optical Solitons in Fibers\/},
Springer-Verlag, Berlin, 1989.

\bibitem{HeLi10} J.~He and Y.~Li, \emph{Designable integrability of the
variable coefficient nonlinear Schr\"{o}dinger equations\/},~Studies in
Applied Mathematics \textbf{126 }(2011)~\#3, (15 pages),
doi:10.1111/j.1467-9590. 2010.00495.x.

\bibitem{Heetal09} X-G. He, D.~Zhao, L.~Lee and H-G.~Luo, \emph{Engineering
integrable nonautonomous Schr\"{o}dinger equations\/},~Phys. Rev. E \textbf{%
79} (2009), 056610 (9 pages).

\bibitem{Hirota71} R.~Hirota, \emph{Exact solution of the Korteweg--de Vries
equation for multiple collisions of solitons\/}, Phys. Rev. Lett. \textbf{27}
(1971)~\#18, 1192--1194.

\bibitem{HirotaBook} R.~Hirota, \textsl{The Direct Method in Soliton Theory\/%
}, Cambridge University Press, Cambridge, 2004.

\bibitem{Kruglovetal03} Y.~I.~Kruglov, A.~C.~Peacock and J.~D.~Harvey, \emph{%
Exact self-similar solutions of the generalized nonlinear Schr\"{o}dinger
equation with distributed coefficients\/}, Phys. Rev. Lett. \textbf{90}
(2003)~\#11, 113902 (4 pages).

\bibitem{Krugloveta05} Y.~I.~Kruglov, A.~C.~Peacock and J.~D.~Harvey, \emph{%
Exact solutions of the generalized Schr\"{o}dinger equation with distributed
coefficients\/}, Phys. Rev. E \textbf{71} (2005), 1056619 (11 pages).

\bibitem{KudryashovBook10} N.~A.~Kudryashov, \textsl{Methods of Nonlinear
Mathematical Physics\/}, Intellect, Dolgoprudny, 2010 [in Russian].

\bibitem{Kundu09} A.~Kundu, \emph{Integrable nonautonomous Schr\"{o}dinger
equations are equivalent to the standard autonomous equation\/}, Phys. Rev.
E \textbf{79} (2009), 015601(R) (4 pages).

\bibitem{LanLopSus11} N.~Lanfear, R.~M.~Lopez and S.~K.~Suslov, \emph{Exact
wave functions for generalized harmonic oscillators\/}, arXiv:11002.5119v1
[math-ph] 24 Feb 2011.

\bibitem{Lan:Sus} N.~Lanfear and S.~K.~Suslov, \emph{The time-dependent Schr%
\"{o}dinger equation, Riccati equation and Airy functions\/},
arXiv:0903.3608v5 [math-ph] 22 Apr 2009.

\bibitem{Lax68} P.~D.~Lax, \emph{Integrals of nonlinear equations of
evolution and solitary waves}\textit{\/}, Commun. Pure Appl. Math. \textbf{21%
} (1968)~\#5, 467--490.

\bibitem{LiangZhangLiu05} Z.~X.~Liang, Z.~D.~Zhang and W.~M.~Liu, \emph{%
Dynamics of a bright solutions in Bose--Einstein condensates with
time-dependent atomic scattering length in an expulsive parabolic potential\/%
}, Phys. Rev. Lett. \textbf{94} (2005)~\#5, 050402 (4 pages).

\bibitem{Malkin:Man'ko79} I.~A.~Malkin and V.~I.~Man'ko, \textsl{Dynamical
Symmetries and Coherent States of Quantum System\/}, Nauka, Moscow, 1979 [in
Russian].

\bibitem{Malk:Man:Trif73} I.~A.~Malkin, V.~I.~Man'ko and D.~A.~Trifonov,
\emph{Linear adiabatic invariants and coherent states\/}, J. Math. Phys.
\textbf{14} (1973) \#5, 576--582.

\bibitem{Mateev:SalleBook} V.~B.~Matveev and M.~A.~Salle, \textsl{Darboux
Transformation and Solitons\/}, Springer--Verlag, Berlin, 1991.

\bibitem{Me:Co:Su} M.~Meiler, R.~Cordero-Soto, and S.~K.~Suslov, \emph{%
Solution of the Cauchy problem for a time-dependent Schr\"{o}dinger
equation\/}, J. Math. Phys. \textbf{49} (2008) \#7, 072102: 1--27; see also
arXiv: 0711.0559v4 [math-ph] 5 Dec 2007.

\bibitem{Miura68} R.~M.~Miura, \emph{Korteweg-de Vries equation and
generalizations.~I. A Remarkable explicit nonlinear transformation}, J.
Math. Phys. \textbf{9} (1968)~\#9, 1202--1204.

\bibitem{Miuraetal68} R.~M.~Miura, C.~S.~Gardner and M.~D.~Kruskal, \emph{%
Korteweg-de Vries equation and generalizations.~II. Existence of
concervation laws and constants of motion}, J. Math. Phys. \textbf{9}
(1968)~\#9, 1204--1209.

\bibitem{Moores96} J.~D.~Moores, \emph{Nonlinear compression of chirped
solitary waves with and without phase modulation}\textit{\/}, Optics Letters
\textbf{21} (1996)~\#8, 555--557.

\bibitem{Moores01} J.~D.~Moores, \emph{Oscilatory solitons in a novel
integrable model of asynchronomous mode locking\/}, Optics Letters \textbf{21%
} (2001)~\#2, 87--89.

\bibitem{Musette99} M.~Musette, \emph{Painlev\'{e} analysis for nonlinear
partial differential equations\/},~in: \textsl{The Painlev\'{e} Property,
One Century Later\/}, CRM Series in Mathematical Physics (R.~Conte, Ed.),
Springer--Verlag, New York, 1999. pp.~517--572.

\bibitem{MusConte03} M.~Musette and R.~Conte, \emph{Analytic solitary waves
of nonintegrable equations\/},~Physica~D \textbf{181} (2003), 70--79.

\bibitem{Newell78} A.~C.~Newell, \emph{Nonlinear tunnelling\/},
J.~Math.~Phys. \textbf{19} (1978)~\#5, 1126--1133.

\bibitem{Novikovetal} S.~Novikov, S.~V.~Manakov, L.~P.~Pitaevskii and
V.~E.~Zakharov, \textsl{Theory of Solitons: The Inverse Scattering Method\/}%
, Kluwer, Dordrecht, 1984.

\bibitem{OlverPBook} P.~J.~Olver, \textsl{Applications of Lie Group to
Differential Equations\/}, Springer--Verlag, Berlin, 1991.

\bibitem{Oraevsky01} A.~N.~Oraevsky, \emph{Bose condensates from the point
of view of laser physics\/}, Quantum Electronics \textbf{31} (2001)~\#12,
1038--1057.

\bibitem{Per-GTorrKonot06} V.~M.~P\'{e}rez-Garc\'{\i}a, P.~Torres and
V.~V.~Konotop, \emph{Similarity transformations for nonlinear Schr\"{o}%
dinger equations with time-dependent coefficients\/}, Physica~D \textbf{221}
(2006), 31--36.

\bibitem{Pit:StrinBook} L.~Pitaevskii and S.~Stringari, \textsl{%
Bose--Einstein Condensation\/}, Oxford University Press, Oxford, 2003.

\bibitem{PonomAgr07} S.~Ponomarenko and G.~P.~Agrawal, \emph{Optical
similaritons in nonlinear waveguides\/}, Optics Letters \textbf{32}
(2007)~\#12, 1659--1661.

\bibitem{Rogers:SchiefBook} C.~Rogers and W.~K.~Scheif, \textsl{B\"{a}cklund
Transformation and Darboux Transformation\/}, Cambridge University Press,
Cambridge, 2002.

\bibitem{SatYaj74} J.~Satsuma and N.~Yajima, \emph{Initial value problems of
one-dimensional self-modulation of nonlinear waves in dispersive media\/},
Supp. Prog. Theor. Phys. \textbf{85} (1974), 284--306.

\bibitem{Scott03} A.~Scott, \textsl{Nonlinear Science: Emergence and
Dynamics of Coherent Structures\/}, Second Edition, Oxford University Press,
Oxford, 2003.

\bibitem{SegurAb81} H.~Segur and M.~J.~Ablowitz, \emph{Asymptotic solutions
of nonlinear equations and a Painlev\'{e} transcendent\/}, Physica~D \textbf{%
3} (1981)~\#1--2, 165--184.

\bibitem{Serkin:Hasrgawa00} V.~N.~Serkin and A.~Hasegawa, \emph{Novel
soliton solutions of the nonlinear Schr\"{o}dinger equation model\/}, Phys.
Rev. Lett. \textbf{85} (2000)~\#21, 4502--4505.

\bibitem{Serkin:Hasegawa00} V.~N.~Serkin and A.~Hasegawa, \emph{Soliton
management in the nonlinear Schr\"{o}dinger equation model with varying
dispertion, nonlinearity, and gain\/}, JETP Letters \textbf{72} (2000)~\#2,
89--92.

\bibitem{Serkinetal04} V.~N.~Serkin, A.~Hasegawa and T.~L.~Belyeva, \emph{%
Comment on \textquotedblleft Exact self-similar solutions of the generalized
nonlinear Schr\"{o}dinger equation with distributed
coefficients\textquotedblright \/}, Phys. Rev. Lett. \textbf{92}
(2004)~\#19, 199401 (1~page).

\bibitem{Serkinetal07} V.~N.~Serkin, A.~Hasegawa and T.~L.~Belyeva, \emph{%
Nonautonomous solitons in external potentials\/}, Phys. Rev. Lett. \textbf{98%
} (2007)~\#7, 074102 (4~pages).

\bibitem{Serkinetal10} V.~N.~Serkin, A.~Hasegawa and T.~L.~Belyeva, \emph{%
Nonatonomous matter-wave soliton near the Feshbach resonance\/}, Phys. Rev.
A \textbf{81} (2010), 023610 (19~pages).

\bibitem{Suaz:Sus} E.~Suazo and S.~K.~Suslov, \emph{Cauchy problem for Schr%
\"{o}dinger equation with variable quadratic Hamiltonians\/}, under
preparation.

\bibitem{SuazoSuslovSol} E.~Suazo and S.~K.~Suslov, \emph{Soliton-like
solutions for nonlinear Schr\"{o}dinger equation with variable quadratic
Hamiltonians\/}, arXiv:1010.2504v4 [math-ph] 24 Nov 2010.

\bibitem{Suslov10} S.~K.~Suslov, \emph{Dynamical invariants for variable
quadratic Hamiltonians\/}, Physica Scripta \textbf{81} (2010)~\#5, 055006
(11~pp); see also arXiv:1002.0144v6 [math-ph] 11 Mar 2010.

\bibitem{Tajiri83} M.~Tajiri, \emph{Similarity reduction of the one and two
dimensional nonlinear Schr\"{o}dinger equations\/}, J. Phys. Soc. Japan
\textbf{52} (1983)~\#2, 1908--1917.

\bibitem{Takei02} Y.~Takei, \emph{On exact WKB approach to Ablowitz--Segur's
connection problem for the second Painlev\'{e} equation\/}, ANZIAM J.
\textbf{44} (2002), 111--119.

\bibitem{Tao09} T.~Tao, \emph{Why are solitons stable?\/}, Bull. Amer. Math.
Soc. \textbf{46} (2009)~\#1, 1--33.

\bibitem{TappZab71} F.~D.~Tappert and N.~J.~Zabusky, \emph{Gradient-induced
fission of solitons\/}, Phys. Rev. Lett. \textbf{26} (1971)~\#26, 1774--1776.

\bibitem{Trall-Gin:Drake:Lop-Rich:Trall-Herr:Bir} C.~Trallero-Giner,
J.~Drake, V.~Lopez-Richard, C.~Trallero-Herrero and J.~L.~Birman, \emph{%
Bose--Eistein condensates: Analytical methods for the Gross--Pitaevskii
equation\/},~Phys. Lett. A \textbf{354} (2006), 115--118.

\bibitem{Weissetal82} J.~Weiss, M.~Tabor and G.~Carnevalle, \emph{The Painlev%
\'{e} property for partial differential equation\/}, J.~Math. Phys. \textbf{%
24} (1983)~\#3, 522--526.

\bibitem{Wolf81} K.~B.~Wolf, \emph{On time-dependent quadratic Hamiltonians\/%
},~SIAM J.~Appl. Math. \textbf{40} (1981)~\#3, 419--431.

\bibitem{ZYan03a} Z.~Yan, \emph{The new extended Jacobian elliptic function
expansion algorithm and its applications in nonlinear mathematical physics
equations\/}, Computer Physics Communications \textbf{153} (2003), 145--154.

\bibitem{Zyan04} Z.~Yan, \emph{An improved algebra method and its
applications in nonlinear wave equations\/}, Chaos, Solitons \& Fractals
\textbf{21} (2004), 1013--1021.

\bibitem{ZYan10} Z.~Yan, \emph{Exact analytical solutions for the
generalized non-integrable nonlinear Schr\"{o}dinger equation with varying
coefficients\/}, Phys. Lett. A \textbf{374} (2010), 4838--4843.

\bibitem{ZYan:Konotop09} Z.~Yan and V.~V.~Konotop, \emph{Exact solutions to
three-dimensional generalized nonlinear Schr\"{o}dinger equation with
varying potential and nonlinearities\/}, Phys. Rev. E \textbf{80} (2009),
036607 (9 pages).

\bibitem{Yeon:Lee:Um:George:Pandey93} K-H.~Yeon, K-K.~Lee, Ch-I.~Um,
T.~F.~George and L.~N.~Pandey, \emph{Exact quantum theory of a
time-dependent bound Hamiltonian systems\/},~Phys. Rev. A \textbf{48}
(1993)~\#~4, 2716--2720.

\bibitem{Zakh:Shab71} V.~E.~Zakharov and A.~B.~Shabat, \emph{Exact theory of
two-dimensional self-focusing and one-dimensional self-modulation of waves
in nonlinear media\/}, Zh. Eksp. Teor. Fiz. \textbf{61} (1971), 118--134
[Sov. Phys. JETP \textbf{34} (1972)~\#1, 62--69.]

\bibitem{ZakhShab74} V.~E.~Zakharov and A.~B.~Shabat, \emph{A scheme for
integrating the nonlinear equations of mathematical physics by the method of
the inverse scattering problem. I\/}, Functional Analysis and Its
Applications \textbf{8 }(1974)~\#3, 226--235.

\bibitem{ZakhShab79} V.~E.~Zakharov and A.~B.~Shabat, \emph{Integration of
nonlinear equations of mathematical physics by the method of inverse
scattering. II\/}, Functional Analysis and Its Applications \textbf{13 }%
(1979)~\#3, 166--174.

\bibitem{Zhangetal08} X.-F.~Zhang, Q.~Yang, J.-F.~Zhang, X.~Z.~Chen and
W.~M.~Liu, \emph{Controlling soliton interactions in Bose--Einstein
condensates by synchronizing the Feshbach resonance and harmonic trap\/},
Phys. Rev. A \textbf{77} (2008), 023613 (7~pages).
\end{thebibliography}
\end{document}